\documentclass[aps,prl,onecolumn,superscriptaddress]{revtex4}
\usepackage{graphicx}

\begin{document}
\title{$(\pi,0)$ antiferromagnetic spin excitations in superconducting Rb$_{0.82}$Fe$_{1.68}$Se$_2$}
\author{Miaoyin Wang}
\affiliation{Department of Physics and Astronomy, The University of Tennessee, Knoxville,
Tennessee 37996-1200, USA}
\author{Chunhong Li}
\affiliation{Beijing National Laboratory for Condensed Matter Physics and Institute of
Physics, Chinese Academy of Sciences, P. O. Box 603, Beijing 100190, China}
\author{D. L. Abernathy}
\affiliation{Neutron Scattering Science Division, Oak Ridge National Laboratory, Oak
Ridge, Tennessee 37831-6393, USA}
\author{Yu Song}
\affiliation{Department of Physics and Astronomy, The University of Tennessee, Knoxville,
Tennessee 37996-1200, USA}
\author{Scott V. Carr}
\affiliation{Department of Physics and Astronomy, The University of Tennessee, Knoxville,
Tennessee 37996-1200, USA}
\author{Xingye Lu}
\affiliation{Beijing National Laboratory for Condensed Matter Physics and Institute of
Physics, Chinese Academy of Sciences, P. O. Box 603, Beijing 100190, China}
\affiliation{Department of Physics and Astronomy, The University of Tennessee, Knoxville,
Tennessee 37996-1200, USA}
\author{Shiliang Li}
\affiliation{Beijing National Laboratory for Condensed Matter Physics and Institute of
Physics, Chinese Academy of Sciences, P. O. Box 603, Beijing 100190, China}     
\author{Jiangping Hu}
\affiliation{Department of Physics, Purdue University, West Lafayette, Indiana 47907, USA}
\affiliation{Beijing National Laboratory for Condensed Matter Physics and Institute of
Physics, Chinese Academy of Sciences, P. O. Box 603, Beijing 100190, China}   
\author{Tao Xiang}
\affiliation{Beijing National Laboratory for Condensed Matter Physics and Institute of
Physics, Chinese Academy of Sciences, P. O. Box 603, Beijing 100190, China}   
\author{Pengcheng Dai}
\email{pdai@utk.edu}
\affiliation{Department of Physics and Astronomy, The University of Tennessee, Knoxville,
Tennessee 37996-1200, USA}
\affiliation{Beijing National Laboratory for Condensed Matter Physics and Institute of
Physics, Chinese Academy of Sciences, P. O. Box 603, Beijing 100190, China}


\begin{abstract}
{
We use inelastic neutron scattering to show that superconducting (SC) rubidium iron selenide 
Rb$_{0.82}$Fe$_{1.68}$Se$_2$ exhibits antiferromagnetic (AF) spin excitations near the in-plane 
wave vector $Q=(\pi,0)$ identical to that for iron arsenide superconductors.  Moreover, we find that
these excitations change from incommensurate to commensurate with increasing energy, and  
occur at the expense of spin waves associated with the coexisting $\sqrt{5}\times\sqrt{5}$ block AF phase.  Since angle resolved photoemission experiments reveal no evidence for hole-like Fermi surface at $\Gamma(0,0)$, our results suggest 
that the $Q=(\pi,0)$ excitations in SC Rb$_{0.82}$Fe$_{1.68}$Se$_2$ come from localized moments and may have a similar origin as the hourglass-like spin excitations in copper oxide superconductors.
}
\end{abstract}

\maketitle

{\bf Introduction} The family of alkaline iron selenide superconductors  
$A_y$Fe$_{1.6+x}$Se$_2$ ($A=$ K, Rb, Cs) \cite{jgguo,krzton,mhfang,afwang} has generated considerable 
interest because superconductivity in these materials may have a different origin from 
the sign reversed $s$-wave electron pairing mechanism \cite{yzhang,tqian,Dxmou}, a leading candidate for superconductivity in iron pnictide superconductors \cite{mazin2011n}.
Although $A_y$Fe$_{1.6+x}$Se$_2$ materials are isostructural with the metallic iron pnictides such as (Ba,Ca,Sr)Fe$_2$As$_2$ \cite{johnston}, they are insulators near $x=0$ \cite{mhfang,afwang} 
and form a $\sqrt{5}\times\sqrt{5}$ block AF structure with Fe vacancy order 
(Fig. 1a)  \cite{haggstrom,wbao1,pomjakushin1,yfeng,mwang11}
completely different from the collinear AF structure of the iron pnictides  \cite{cruz}.  Since 
superconductivity in $A_y$Fe$_{1.6+x}$Se$_2$ always appears concurrently 
with the block AF order \cite{wbao1,pomjakushin1,yfeng,mwang11}, whereas in the iron pnictides  superconductivity arises at the 
expense of the static AF order \cite{cruz}, 
it is important to determine the relationship between superconductivity and magnetism in these materials.
Although experiments using transmission electron microscopy \cite{yjsong}, X-ray diffraction \cite{ricci}, muon-spin rotation ($\mu$SR) \cite{shermadini}, scanning tunneling microscopy \cite{wli}, angle resolved photoemission (ARPES) \cite{fchen},
M${\rm \ddot{o}}$ssbauer \cite{ksenofontov}, and optical \cite{charnukha,homes} spectroscopy have provided tantalizing evidence for several coexisting phases in superconducting (SC) $A$Fe$_{1.6+x}$Se$_2$, it is still 
unclear what is the exact crystal structure and stoichiometry of the SC phase and its relationship to the
$\sqrt{5}\times\sqrt{5}$ AF phase.

For iron pnictides \cite{johnston}, band structure calculations have predicted that 
Fermi surfaces of these materials are composed of hole and electron pockets near $\Gamma(0,0)$ and $M(\pi,0)/M(0,\pi)$ points, respectively \cite{mazin2011n}.  Since antiferromagnetism and 
superconductivity can arise from the sign reversed   
quasiparticle excitations between the hole and electron pockets \cite{mazin2011n},
there should be a neutron spin resonance at the in-plane wave vector $Q=(\pi,0)$ \cite{maier,korshunov}.
Indeed, inelastic neutron scattering experiments on single 
crystals of electron and hole-doped BaFe$_2$As$_2$  
have found the resonance at $Q=(\pi,0)$ \cite{lumsden,chi,inosov,clzhang} and thus provided evidence for 
the electron $s^\pm$-wave pairing mechanism \cite{mazin2011n}.  In the case of SC $A_y$Fe$_{1.6+x}$Se$_2$,
since ARPES measurements \cite{yzhang,tqian,Dxmou} found electron Fermi surfaces at the $M(\pi,0)/M(0,\pi)$ points but no hole Fermi surface 
near $\Gamma(0,0)$, quasiparticle excitations between   $\Gamma(0,0)$ and $M(\pi,0)/M(0,\pi)$ should not provide AF spin excitations at $Q=(\pi,0)$ (Fig. 1c).  Instead,    
the nesting properties between the $M(\pi,0)/M(0,\pi)$ electron pockets in a $d$-wave symmetry is expected to give a broad plateau like maximum around $Q=(\pi,\pi)$ that is bordered by two peaks at $Q\approx (\pi,0.625\pi)$
and $Q\approx (0.625\pi,\pi)$ \cite{maier11}.  Although the recent discovery of the neutron spin resonance in 
SC Rb$_y$Fe$_{1.6+x}$Se$_2$ at wave vectors $Q=(\pm\pi,\pm0.5\pi)$ [or 
$Q=(\pm0.5\pi,\pm\pi)$] (Fig. 1d) \cite{friemel,park} is consistent with this picture \cite{maier11}, it remains unknown whether there are spin excitations 
at other wave vectors not associated with the Fermi surface nesting.

In this Letter, we use neutron scattering to map out the low-energy spin excitations in SC 
Rb$_{0.82}$Fe$_{1.68}$Se$_2$ ($T_c=32$ K, Fig. 1f).  In addition to confirming the neutron spin resonance at $Q=(\pm\pi,\pm0.5\pi)$ \cite{friemel,park}, we find clear evidence for incommensurate spin excitations near wave vector
$Q=(\pi,0)$ that are absent in 
insulating Rb$_{0.89}$Fe$_{1.58}$Se$_2$ (Figs. 1b and 1d) \cite{mywang11}. With increasing energy, the incommensurate spin excitions disperse inward to
$Q=(\pi,0)$ and disappear above $E=30$ meV (Figs. 2,3).  A comparison of spin excitations in
SC Rb$_{0.82}$Fe$_{1.68}$Se$_2$ with spin waves in insulating    
Rb$_{0.89}$Fe$_{1.58}$Se$_2$ \cite{mywang11} reveals that the intensity gain of the $Q=(\pi,0)$ excitations 
is at the expense of spin waves associated with the 
$\sqrt{5}\times\sqrt{5}$ AF phase (Fig. 3).  Since electron-hole pocket excitations are impossible between $\Gamma(0,0)$ and $M(\pi,0)/M(0,\pi)$ points \cite{yzhang,tqian,Dxmou}, our results suggest the presence of local moments \cite{qmsi} in addition to the itinerant electron induced resonance \cite{friemel,park}.
Moreover, the dispersion of the $Q=(\pi,0)$ excitations is similar to that of copper oxide 
superconductors \cite{hayden,tranquada} and insulating cobalt oxide \cite{boothroyd}, thus suggesting
the possible presence of dynamic stripes \cite{kivelson}.

\begin{figure}[t]
\includegraphics[scale=.4]{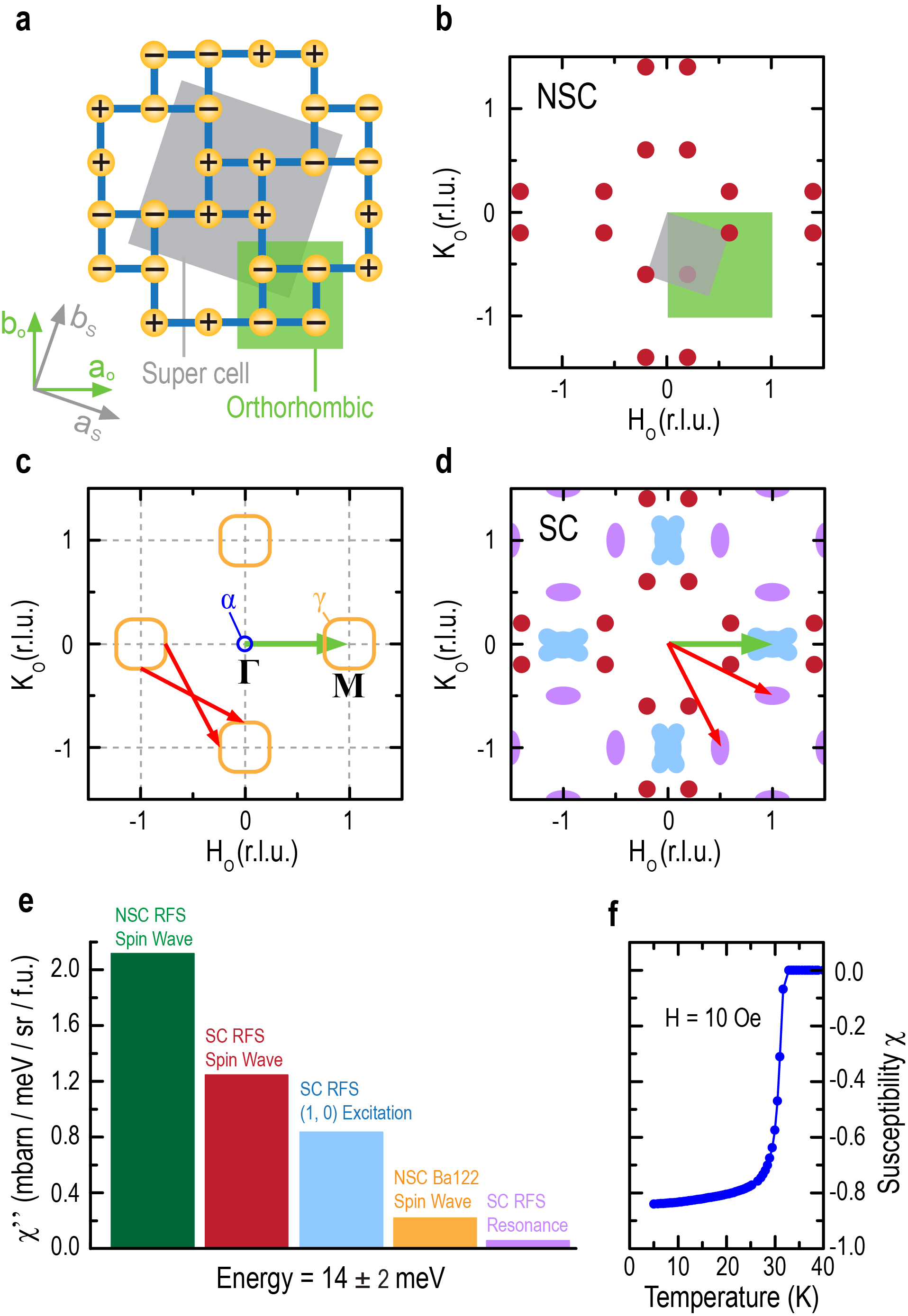}
\caption{(Color online) (a) The block antiferromagnetic spin structure of the 
insulating $A_y$Fe$_{1.6+x}$Se$_2$, where 
the $\sqrt{5}\times\sqrt{5}$ superlattice structure is marked as grey with lattice parameter $a_s=8.663$ \AA\ and the orthorhombic lattice cell similar to iron pnictides is shaded green \cite{mywang11}.
(b) The reciprocal space in the $[H_o,K_o]$ plane, 
where the solid red circles indicate the AF Bragg peak positions.
(c)	Schematics of the Fermi surfaces of SC $A_y$Fe$_{1.6+x}$Se$_2$
from ARPES measurements.  There are four large electron pockets at $Q=(\pm1,0)/(0,\pm1)$ and a small
electron pocket at $\Gamma(0,0)$ \cite{yzhang,tqian,Dxmou}.  The neutron spin resonance is believed to 
originate from the electron-electron pocket excitations as shown by the red 
arrows \cite{friemel,park}. The green arrow indicates the $\Gamma\leftrightarrow M$ transition.
(d) Positions of observed spin excitations in SC Rb$_{0.82}$Fe$_{1.68}$Se$_2$, where spin waves from
the block AF phase, neutron spin resonance, and $(\pi,0)$ excitations are marked as 
red solid circles, purple ellipses and light-blue cross shapes, respectively.
(e) Integrated intensity comparison of several samples at $E=14\pm 2$ meV. Olive Green: spin waves in 
the insulating Rb$_{0.89}$Fe$_{1.58}$Se$_2$; Dark red, light blue, and light violet are spin waves,
$(\pi,0)$ excitations, and resonance in SC 
Rb$_{0.82}$Fe$_{1.68}$Se$_2$; Orange: spin wave in BaFe$_2$As$_2$. (f) Susceptibility measurement indicates $T_c=32$ K.
}
\end{figure}

\begin{figure}[t]
\includegraphics[scale=.45]{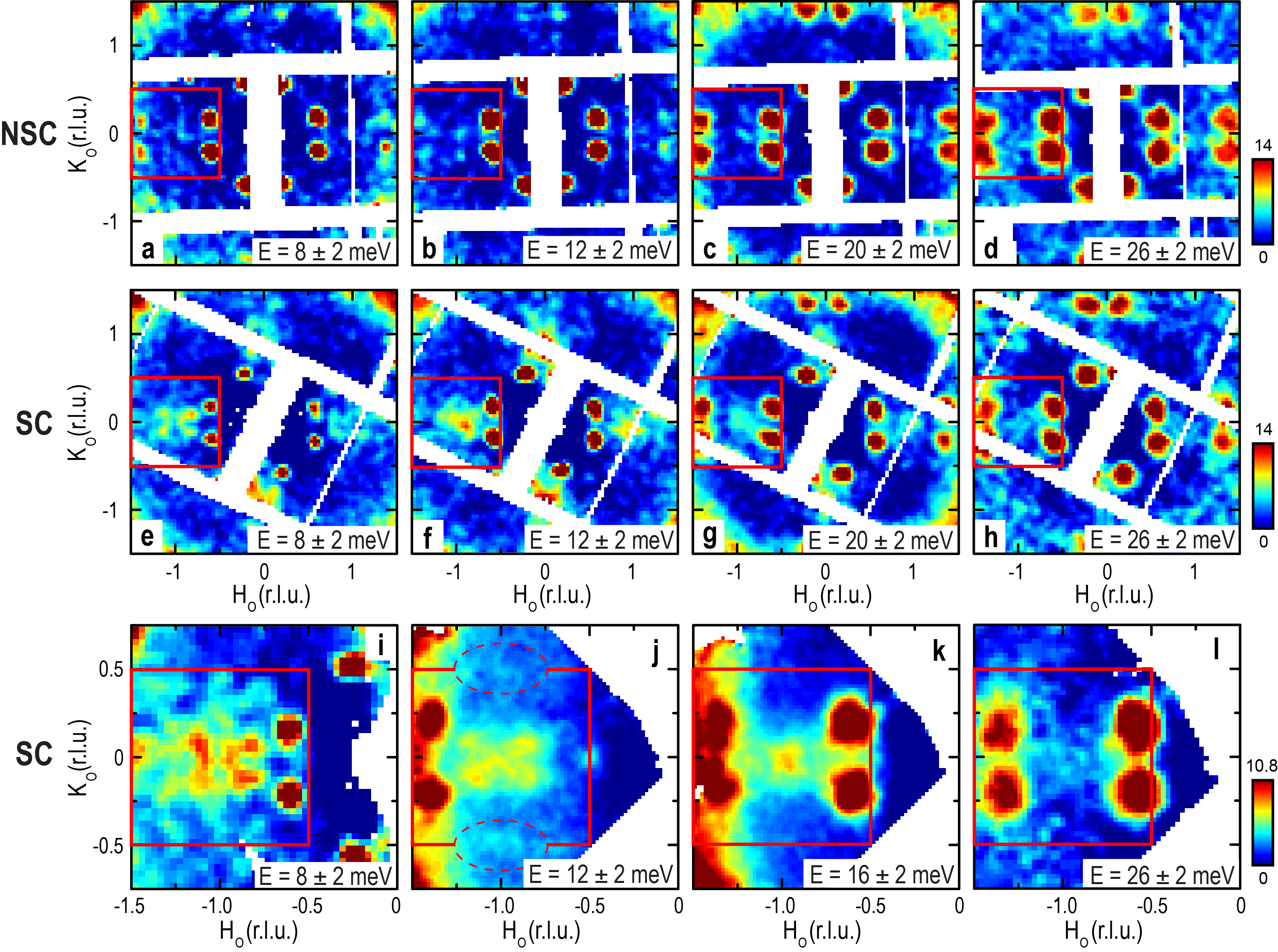}
\caption{(Color online) (a-d) Wave-vector dependence of spin-wave excitations at different energies 
for NSC Rb$_{0.89}$Fe$_{1.58}$Se$_2$ at 10 K obtained with 
incident neutron energy of $E_i=80$ meV \cite{mywang11}.  
(e-h) Identical images for SC Rb$_{0.82}$Fe$_{1.68}$Se$_2$ at 6 K. The red squares are the Brillouin zone for iron pnictides \cite{harriger}.
(i-l) Expanded view of the excitations near $Q=(1,0)$.  The data in (j,k) are collected with $E_i=35$ meV, while
(i,l) are taken with $E_i=80$ meV.  The dashed ellipses in (j) mark positions of the resonance.
The vertical color bars indicate intensity scale in mbarns/sr/meV/f.u.
}
\end{figure}

{\bf Results} We have performed inelastic neutron scattering experiments on the ARCS chopper spectrometer at the Spallation Neutron Source, Oak Ridge National Laboratory using identical conditions 
as previous work on spin waves in insulating Rb$_{0.89}$Fe$_{1.58}$Se$_2$ \cite{mywang11}.
Figures 1a and 1b show the $\sqrt{5}\times\sqrt{5}$ block AF structure and the positions 
of the AF peaks in reciprocal space, respectively \cite{mywang11}. 
We define the wave vector $Q$ at $(q_x, q_y, q_z)$ as 
$(H_o,K_o,L_o)=(q_xa_o/2\pi,q_ya_o/2\pi,q_zc_o/2\pi)$ rlu,
where $a_o=5.48$ and $c_o=14.69$ \AA\ are the orthorhombic cell lattice parameters similar 
to iron pnictides \cite{harriger}.  In this notation, 
the neutron spin resonance \cite{friemel,park} occurs at 
$Q=(\pm1,\pm0.5)$ [or $Q=(\pm\pi,\pm0.5\pi)$] (Fig. 1d), while 
the $\Gamma\leftrightarrow M$ Fermi surface nesting gives scattering at $Q=(\pm1,0)$ rlu (Figs. 1c and 1d).
We co-aligned $\sim$6 grams of the SC 
single crystals Rb$_{0.82}$Fe$_{1.68}$Se$_2$ grown by self-flux method 
(with mosaic of $\sim$6$^\circ$) \cite{mywang11}, where  
the chemical composition was determined by
inductively-coupled plasma analysis.  
Figure 1f shows the temperature dependence of the susceptibility measurements confirming 
 $T_c = 32$ K. To ensure that the neutron 
 spin resonance at $Q=(-1,0.5)$ at $E=14$ meV \cite{friemel,park} does not fall into detector gaps on ARCS,
we rotated the co-aligned samples counter-clockwise by $\sim$27 degrees. 
The incident beam energies were $E_i = 35, 80$ meV with $E_i$ parallel to 
the $c$-axis. The scattering intensities were normalized
to absolute units using a vanadium standard and can therefore be compared directly with spin waves in insulating Rb$_{0.89}$Fe$_{1.58}$Se$_2$ \cite{mywang11}.

\begin{figure}[t]
\includegraphics[scale=.4]{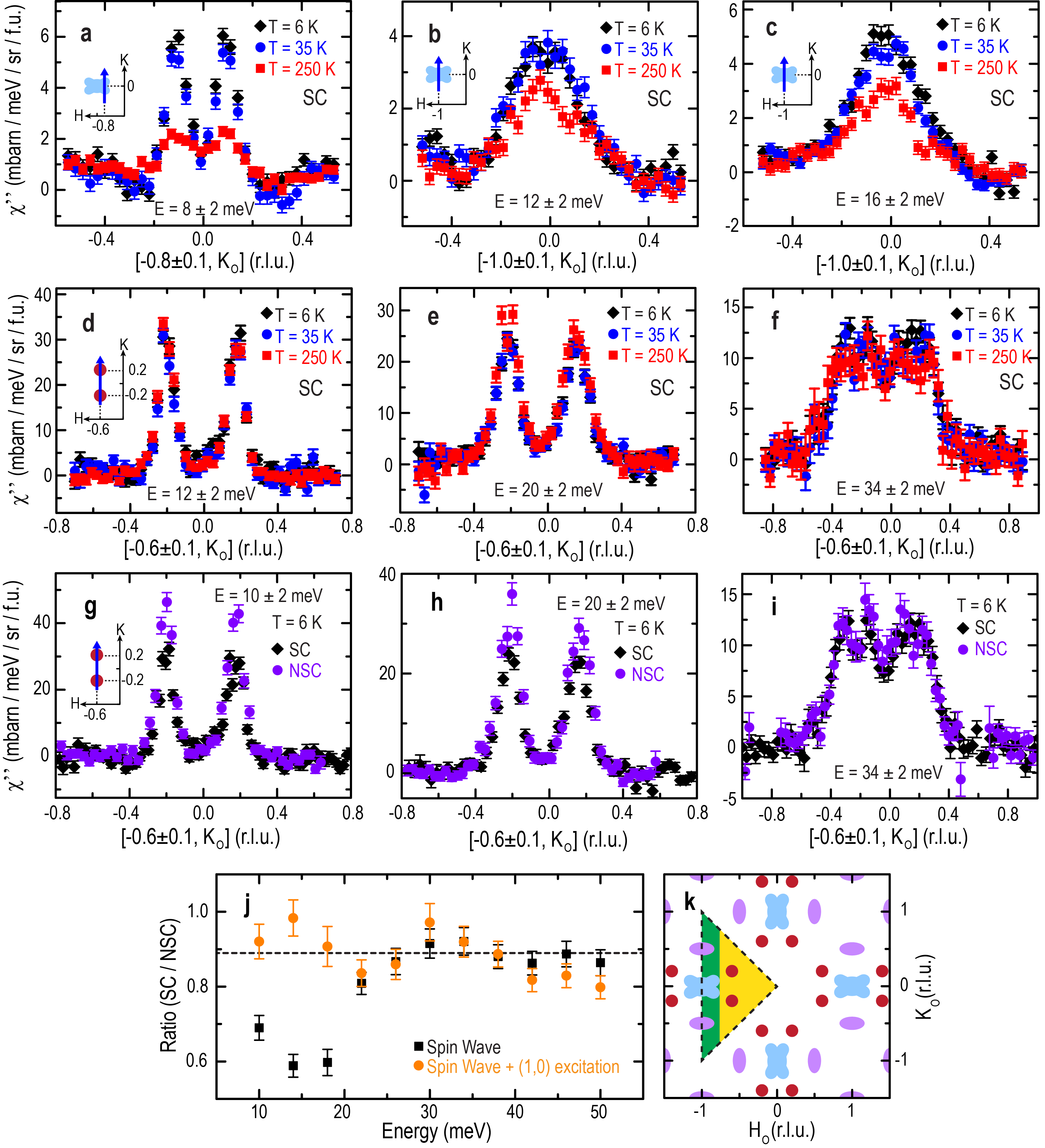}
\caption{ 
Cuts of $\chi^{\prime\prime}(Q,\omega)$ along (a) the $[-0.8\pm0.1,K]$, (b,c) $[-1\pm0.1,K]$ directions for the 
$Q=(-1,0)$ excitations at different temperatures. 
(d-f) Cuts of spin waves along the
$[-0.6\pm0.1,K]$ direction at different energies and temperatures reveal that
$\chi^{\prime\prime}(Q,\omega)$ is temperature independent up to $T=250$ K.
(g-i) Comparison of the low-temperature 
spin wave intensities for SC and insulating samples using
the same cuts along the $[-0.6\pm0.1, K]$ direction. The spin wave intensity of the SC sample are
lower at $E=12\pm 2$ and $20\pm 2$ meV but become similar as that of 
the insulating sample at $E=34\pm 2$ meV.
(j) 
The black squares are ratio of spin waves in yellow area for SC and insulating samples. 
The yellow circles are the ratio of excitations in yellow area + green area for SC and insulating samples. 
}
\end{figure}

From earlier work on $A_y$Fe$_{1.6+x}$Se$_2$
 \cite{wbao1,pomjakushin1,yfeng,mwang11}, we know that superconductivity coexists with  
the block AF order.  Therefore, one should expect acoustic spin waves 
in SC Rb$_{0.82}$Fe$_{1.68}$Se$_2$ from the block AF phase \cite{mywang11}.
Figure 2 summarizes the two-dimensional constant-energy 
($E$) images of spin excitations in the $[H_o,K_o]$ plane
for insulating and SC Rb$_y$Fe$_{1.6+x}$Se$_2$.
Since the subtle changes in the insulating
and SC samples \cite{wbao1,pomjakushin1,yfeng,mwang11} are not expected to much affect phonons
in these materials, we assume 
that the new dispersive features in Rb$_{0.82}$Fe$_{1.68}$Se$_2$ are spin excitations
associated with the SC phase.  Figures 2a-2d show images of acoustic spin waves  
at energies $E=8\pm2$, $12\pm2$, $20\pm2$, and $26\pm2$ meV, respectively, 
 for insulating Rb$_{0.89}$Fe$_{1.58}$Se$_2$ \cite{mywang11}.
They are centered at the expected in-plane AF wave vectors  with no
observable features at $Q=(1,\pm0.5)$ and $Q=(1,0)$ \cite{mywang11}.

\begin{figure}[t]
\includegraphics[scale=.4]{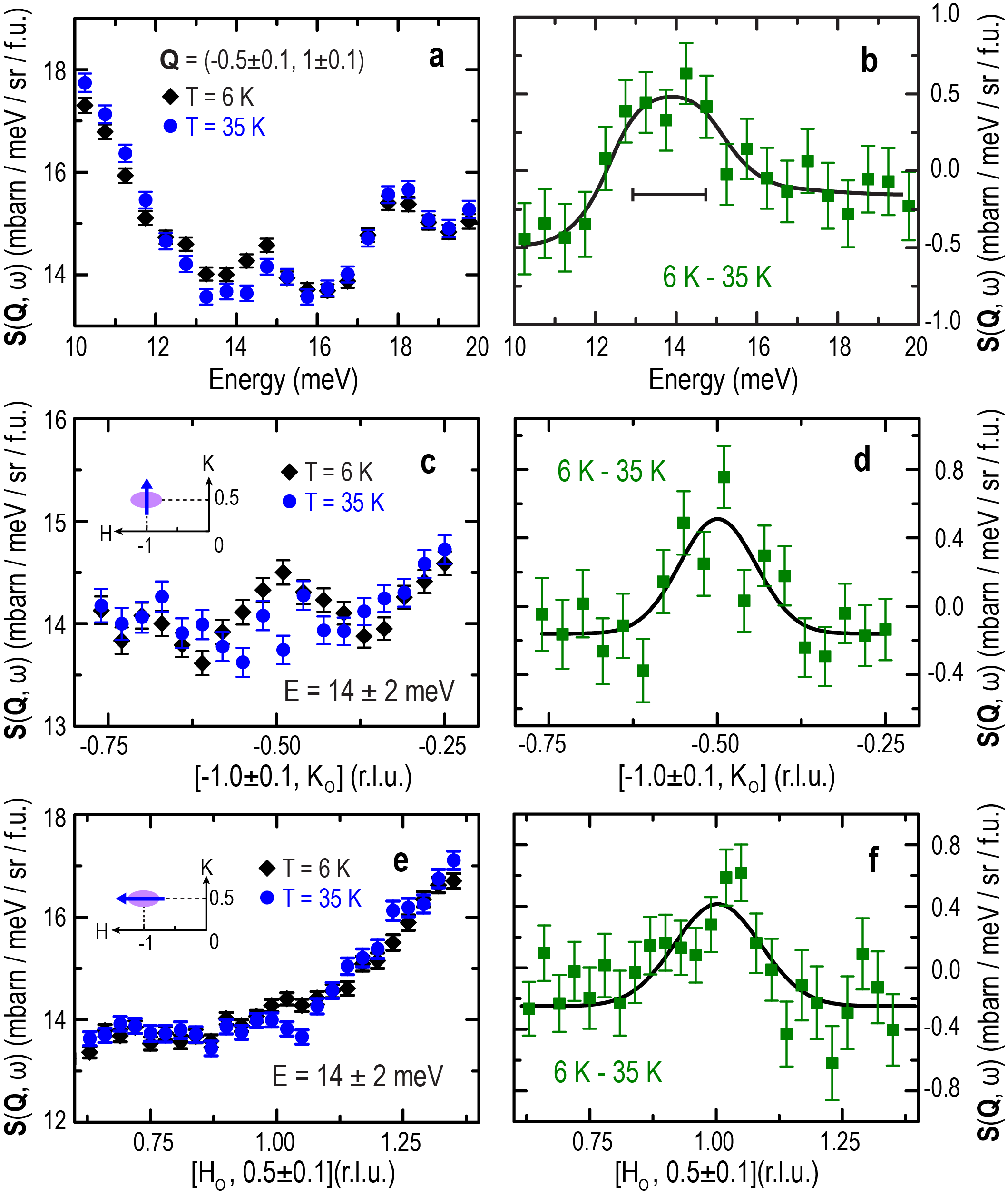}
\caption{
(a) Energy cut at the resonance position by integrating $Q=(-0.5\pm0.1,1\pm0.1)$. 
(b)	Subtracting 35 K data from 6 K data shows a resonance at $E=14$ meV. 
The horizontal bar is the instrumental energy resolution. 
Constant-energy cuts along the (c) $[H,1\pm 0.1]$ and (e)
$[0.5\pm 0.1, K]$ directions. The 6 K$-$35 K data confirm
 the resonance peak at $(1,-0.5)$ with a width $FWHM=0.13\pm0.04$ along the
 $H$ direction and $FWHM=0.20\pm0.05$ along the $K$ direction.
}
\end{figure}

Figures 2e-2h plot images of the identical constant-energy cuts for SC   
Rb$_{0.82}$Fe$_{1.68}$Se$_2$ at $T=6$ K.  In addition to the usual spin waves from the
block AF structure, we find new features near $Q=(\pm1,0)$ and $Q=(0,\pm1)$.
At $E=8\pm 2$ meV, there are four incommensurate peaks centered at  $Q\approx(-1\pm 0.14,\pm 0.1)$ (Fig. 2e).
Upon increasing energies to 
$E=12\pm 2$ (Fig. 2f) and $20\pm 2$ meV (Fig. 2g), the excitations become approximately centered
at  $Q=(\pm1,0)$. Finally at $E=26\pm 2$ meV,
they disappear at  $Q=(\pm1,0)$ and spin waves in SC Rb$_{0.82}$Fe$_{1.68}$Se$_2$
and insulating Rb$_{0.89}$Fe$_{1.58}$Se$_2$ become indistinguishable (Figs. 2d and 2h).   
Figures 2i-2l show the expanded view of the 
spin excitations near $Q=(-1,0)$ at different energies.
At $E=8\pm 2$ meV, we see four distinct peaks (Fig. 2i).  At the neutron spin resonance energy of
$E=12\pm 2$ meV, the excitations become cross-like near $Q=(-1,0)$ 
and one can also see the resonance centered at $Q=(-1,\pm 0.5)$ ( Fig. 2j) \cite{friemel,park}. Upon increasing energy to $E=16\pm 2$ meV, the excitations are well centered at $Q=(-1,0)$ (Fig. 2k).  Finally at $E=26\pm 2$ meV, we find only spin waves 
from the block AF phase centered around the expected AF positions.

To see how the excitations near $Q=(1,0)$ respond to 
superconductivity and determine whether they are related 
to spin waves from the block AF phase, we show in Fig. 3 
constant-energy cuts for the $Q=(1,0)$ excitations and block AF spin waves at different temperatures.
The neutron scattering cross section $S(Q,E)$ is related to the imaginary part of the dynamic susceptibility $\chi^{\prime\prime}(Q,\omega)$ by correcting for the Bose population factor
via $S(Q,E)= 1/(1-\exp(-E/(k_BT)))\chi^{\prime\prime}(Q,E)$, where $k_B$ is the 
Boltzmann's constant.   Figures 3a-3c
show constant-energy cuts along the $K_o$ direction for different temperatures
at $E=8\pm 2$, $12\pm 2$, and $16\pm 2$ meV, respectively.  While $\chi^{\prime\prime}(Q,\omega)$
at the probed energies 
show no appreciable changes across $T_c$, it decreases on warming to $T=250$ K, consistent with
spin excitations.  For comparison, we find that $\chi^{\prime\prime}(Q,\omega)$ of the spin waves from the block AF phase are temperature independent between 10 K and 250 K (Figs. 3d-3f).  This is
expected since spin waves are bosons and should follow the Bose factor below $T_N$.
To see if superconductivity has any effect on spin waves of the block AF phase, we show in 
Figs. 3g-3i $\chi^{\prime\prime}(Q,\omega)$  
for SC Rb$_{0.82}$Fe$_{1.68}$Se$_2$ and insulating Rb$_{0.89}$Fe$_{1.58}$Se$_2$.
While the spin wave intensity at $E=10\pm2$ and $20\pm2$ meV in the superconductor  
are lower than that of the insulator, it becomes similar at $E=34\pm 2$ meV.  
To quantitatively compare the differences between the intensity gain
near $(-1,0)$ with intensity loss of the AF spin waves in superconductor 
compared with that of the insulator, we plot in in Fig. 3j the ratio of yellow
area and yellow plus green areas for SC and insulating samples (Fig. 3k) as
black square and yellow cirlces, respectively.  We see that the spin wave intensity loss
below $\sim$30 meV is approximately compensated by an intensity gain from excitations around $(-1,0)$.

Finally, to confirm the neutron spin resonance near $E=14$ meV at $Q=(-1,0.5)$  in
our SC Rb$_{0.82}$Fe$_{1.68}$Se$_2$ \cite{friemel,park}, we carried out constant-$Q$ and
constant-energy cuts to the data in Fig. 2j below and above $T_c$.  Figure 4a shows the $S(Q,E)$ for 
integrated wave vectors $Q=(-0.5\pm 0.1,1\pm 0.1)$ at 6 K and 35 K.  The temperature difference
plot (6 K$-$35 K)in Fig. 4b has a clear peak at $E=14$ meV, thus confirming the neutron spin resonance in the
SC state \cite{friemel,park}. Figures 2c and 2e show constant-energy cuts along the two different 
high symmetry directions (see insets) below and above $T_c$.  The temperature difference plots show well-defined peaks at the expected wave vector, again consistent with 
previous work \cite{friemel,park}.  Figure 1e compares the strength of the
spin waves from the block AF structure in insulating and SC 
samples, the $(1,0)$ spin excitations, the resonance, and spin waves of 
BaFe$_2$As$_2$ \cite{harriger} near $E=14$ meV.

{\bf Discussion} The discovery of spin excitations near the $(\pi,0)$ AF wave vector and their dispersion in SC 
Rb$_{0.82}$Fe$_{1.68}$Se$_2$ have several important implications. First, 
since ARPES experiments reveal that SC $A_y$Fe$_{1.6+x}$Se$_2$ have no hole-like 
Fermi surface at $\Gamma(0,0)$ \cite{yzhang,tqian,Dxmou}, 
the $(\pi,0)$ spin excitations cannot arise from quasiparticle excitations between $\Gamma$ 
and $M$ points and most likely come from localized magnetic moments \cite{qmsi}.  
Taking into account that SC 
Rb$_{0.82}$Fe$_{1.68}$Se$_2$ also has a neutron spin resonance most likely arising 
from Fermi surface nesting and itinerant electrons \cite{friemel,park}, these results suggest that localized moments and itinerant electrons are both important ingredients for magnetism in alkaline iron selenide superconductors.
Second, the observation of low-energy incommensurate spin excitations and its inverse dispersion are  
reminiscent of the spin excitations for copper oxide superconductors \cite{hayden,tranquada} and insulating 
La$_{2-x}$Sr$_x$CoO$_4$ \cite{boothroyd}.  This suggests that the $(\pi,0)$ spin excitations stem from strongly correlated electronic physics and may be associated with dynamic  
stripes \cite{kivelson}.  Third,    
the reduction in the low-energy spin wave intensity for the block AF phase in SC Rb$_{0.82}$Fe$_{1.68}$Se$_2$ and the concurrent appearance of the incommensurate spin excitations near  
$Q=(\pi,0)$ indicate that spin excitations in superconductors are compensated by  
spin waves in the AF block phase.  If the SC phase in 
Rb$_{0.82}$Fe$_{1.68}$Se$_2$ mesoscopically coexists with 
the block AF phase \cite{yjsong,ricci,shermadini,wli,fchen,ksenofontov,charnukha,homes}, 
 one can imagine the formation of a striped phase on the interface region of the block AF phase and the SC phase due to the interaction between local moments and itinerant electrons.  The latter can be viewed as dopants to a Mott insulator phase and natually result in a stripe phase as in the case of copper oxides \cite{kivelson}.

\begin{flushleft}
{\bf Acknowledgements} 
We thank Qimiao Si for helpful discussions and R. Zhang for work at IOP.
The work at UTK is supported by the U.S. NSF-DMR-1063866 and NSF-OISE-0968226. 
Work at IOP is supported by the MOST of China 973 program (2012CB821400) and by NSFC-51002180.
ORNL neutron scattering facilities are supported by
the U.S. DOE, Division of Scientific User Facilities. 
\end{flushleft}


\end{document}